\documentclass[twocolumn,showpacs,amsmath,amssymb,prl,superscriptaddress]{revtex4}
\usepackage{graphicx}

\begin{document}
\date{\today}
\author{Jean-Fran\c{c}ois M\'etayer}
\email{jean-francois.metayer@ds.mpg.de}
\affiliation{Max Planck Institute for Dynamics and Self-Organization, G\"ottingen, Germany }

\author{Donald J. Suntrup III}
\affiliation{Center for Nonlinear Dynamics and Department of Physics, University of Texas at Austin, USA}

\author{Charles Radin} 
\affiliation{Department of Mathematics, University of Texas at Austin, USA}

\author{Harry L. Swinney} 
\affiliation{Center for Nonlinear Dynamics and Department of Physics, University of Texas at Austin, USA}

\author{Matthias Schr\"oter}
\email{matthias.schroeter@ds.mpg.de}
\affiliation{Max Planck Institute for Dynamics and Self-Organization, G\"ottingen, Germany }

\title{Shearing of frictional sphere packings}

\begin{abstract}
We measure shear response in packings of glass beads by pulling a thin, 
rough, metal plate vertically through a bed of volume fraction $\phi$, which is
set, before the plate is pulled, in the range 0.575 to 0.628.
The yield stress is  velocity independent over 4 decades and  
increases exponentially with $\phi$, with a transition at $\phi \approx 0.595$.
An analysis of the measured force fluctuations indicates that the shear modulus 
is significantly smaller than the bulk modulus.
\end{abstract}

\maketitle

\section{Introduction}

Granular matter is distinguished in part by its
response to shear stress. We examine here the response to shear of beds
of glass beads of diameter 200 $\mu$m. The granular bed is
prepared (by fluidization
and sedimentation in water) in a well-defined homogeneous 
volume fraction $\phi$ in the
range $0.575-0.628$ . We measure the dependence on volume
fraction of the response of the bed to shear by pulling a thin, rough, 
metal plate embedded in
the material. We find a transition indicated by
a change of slope of yield force as a function of $\phi$ at
$\phi\approx 0.595$, consistent with the phase transition at 
$\phi$ between  $0.59 $ and $0.60$
reported in \cite{schroeter:07}. In the Discussion Section 
we argue that the transition in the yield
slope can be identified with the dilatancy onset, the 
change in sign of the volume response of a sheared system. Further,
we suggest that the 
yielding itself can be viewed as a separate transition 
between a solid-like mechanically stable state and a granular fluid state. 
In recent years the latter transition has often been discussed in the context of the 
Jamming paradigm \cite{van_hecke:10} proposed by Liu and Nagel \cite{liu:98}.  
We show in the Discussion section how the yield stress can be
used to distinguish the regimes in a jamming state diagram \cite{liu:98,ohern:03} for frictional
particles.

\section{Experiment}

\begin{figure}
  \centering
   \includegraphics[width=8.2cm,trim=220 0 0 0]{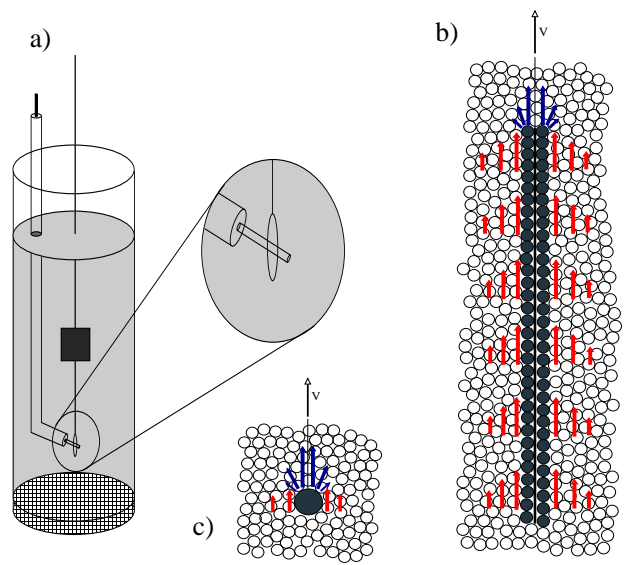}
  \caption{(a) Sketch of the experimental setup. The enlargement shows how the plate and nylon thread are held
in place during fluidization; then the force required to withdraw the plate is measured.
(b) Cartoon of the sheared zone (red) on the sides of the plate and the compressed zone (blue) on top of it.
(c) A rod pulled through the bed experiences a much smaller shear component.}
  \label{fig:setup}
\end{figure}

Our experimental set-up (Fig. \ref{fig:setup}(a)) consists of a vertical glass tube (diameter 38.7 mm, height 550 mm)
filled with water and soda-lime glass beads (diameter $d$ = 200$\pm$10 $\mu$m, average sample height 300 mm).  
The granular bed is fluidized by 
a computer-controlled gear pump that produces 1 min long flow pulses,
which are followed by a 2 min settling time
during which the bed 
completely sediments \cite{schroeter:07}.  This yields a homogeneous bed with a packing 
fraction $\phi$ known with an accuracy of $10^{-4}$. 
The packing fraction after sedimentation depends on the flow rate 
during fluidization: loose packings are obtained using  a small number of pulses of flow rates 
on the order of 50 ml/min, while larger packing fractions are obtained by fluidizing 
the granular bed repeatedly with pulses of decreasing flow rates. 
To check for the role of the preparation history we also performed some measurements where
the bed was fluidized with a syringe pump.

To measure the shear force needed to initiate a displacement in the granular bed, a thin metallic plate (12.8 
mm high $\times$ 9.8 mm wide $\times$ 0.1 mm thick)  is pulled vertically through the bed
(depicted in Fig. \ref{fig:setup}(b). Both sides of the
plate are covered with glass beads (glued there with 3M
Bonding film 583); the resulting total thickness of the plate is  0.6 mm. 
The measured vertical force results both from the shear at the sides of the plate and from  compression 
due to the upward motion of the top edge of the plate. To determine the value of the compression 
component we made measurements using a horizontal cylindrical rod that has the same length as the plate width ($9.8$mm) and a 
diameter the same as the thickness of the plate ($0.6$mm) (Fig. \ref{fig:setup}(c)).

The plate is immersed in the bed before fluidization and is connected to a load cell 
(Honeywell Sensotec Model 31, range 250 g) 
by a nylon thread (diameter 150 $\mu$m, length 275 mm, elastic constant 10 N/mm).
To keep the plate in place during the fluidization, a nylon thread is attached to the plate bottom
and the other end of the thread is attached to a plastic rod (cf. 
Fig. \ref{fig:setup}). 
After the sedimentation following the fluidization pulses, the plastic rod is withdrawn through a metallic cylinder
 to minimize disturbances to the granular bed. Then the plate is  free to move. 

During a measurement, a translation stage driven by a DC-Servo motor pulls the plate upwards 
over a distance of 1  mm at a constant
velocity of 0.15 mm/min (except for a set of measurements made to determine the velocity dependence).
Before the measurement the upper edge of the plate is 160 $\pm$ 5 mm below the sand surface.
The shear force is measured with the load cell connected to a bridge amplifier (Omega DMD-465WB); its
output voltage is digitized with a 16 bit data acquisition card (NI PCI-6036E) at 
a frequency of  100 Hz.  This corresponds to one data point every
$1 \times 10^{-4}$  grain diameters. 
The noise level of this setup corresponds to about 1 mN, as determined from a measurement lifting 
a plate in an empty tube. 

\section{Results}
\begin{figure}
  \includegraphics*[width=9.8cm]{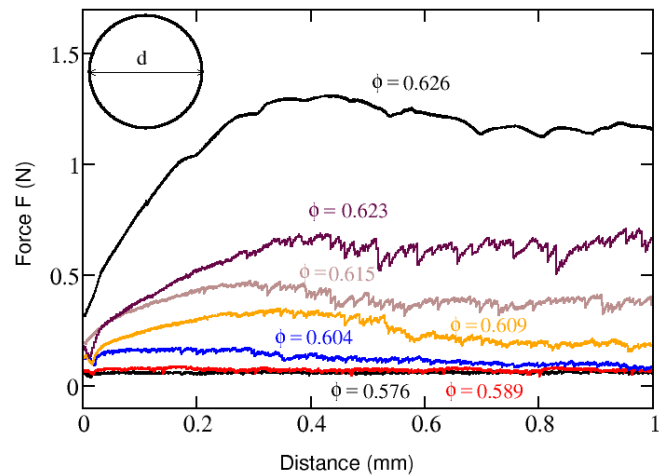}
  \caption{Force needed to  pull the plate up at different packing fractions.
  The circle corresponds to one bead diameter.}
  \label{fig:stress_displacement}
\end{figure}

\begin{figure}
  \includegraphics*[width=8.5cm]{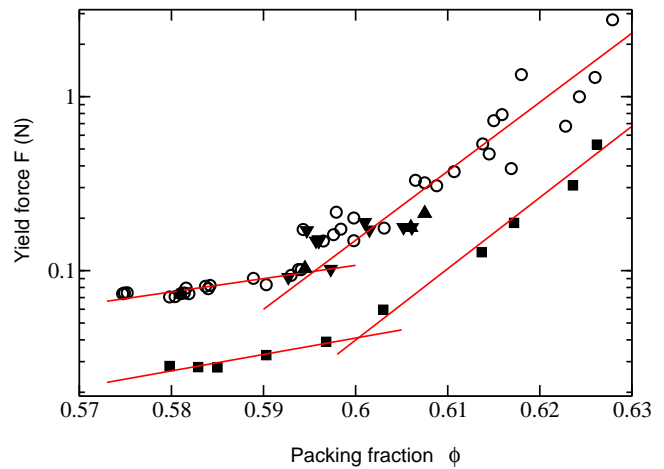}
  \caption{Yield force as a function of the volume fraction. Open circles and filled triangles correspond to the force required to pull
the plate. Filled squares show the force needed to pull a horizontal rod of the same horizontal dimensions as the upper edge of the plate.
The solid lines are exponential fits that intersect at $0.595$ for the plate and $0.6$ for the rod.  
The beds were prepared using a gear pump (circles and squares), a syringe pump 
delivering a sequence of flow pulses with decreasing flow rates (up triangles), and a syringe pump delivering a large number of constant small flow rate 
pulses (down triangles). The datum at the highest volume fraction was taken with a Model 31 load cell 
with the range 10 N.}
  \label{fig:yield_phi}
\end{figure}

Typical measurements of the force as a function of the displacement of the plate are presented in 
Fig. \ref{fig:stress_displacement}. For $\phi < 0.59$, the force is constant for the whole displacement. For larger $\phi$ the force reaches 
a maximum value in the first $0.5$ mm before it begins to decrease again. 
We take this maximum to be the yield stress, the force needed to initiate a reorganization of the structure of the bed. 
This yield force is measured as the average over a length of approximately two grain diameters 
centered on the position of the maximum.  
The dependence of the yield force on packing fraction is shown in Fig. \ref{fig:yield_phi} as open circles.
Though the data are noisy,  there is a clear change in slope at $\phi \approx 0.595$.
To test if the noise stems from minute differences in the packing preparation, which are unavoidable using a gear pump, 
we prepared samples using a syringe pump and two different protocols of flow pulses (cf. the triangular points in Fig.~\ref{fig:yield_phi}). 
The absence of a systematic shift of these measurements suggests that the noise is 
due to the finite size of our probe.

As discussed above, the plate not only shears the sample but also exerts compression. To quantify this compression
we measured the yield force when the horizontal rod was pulled vertically (cf. Fig. 1(c)); the results are given by the filled squares in Fig. \ref{fig:yield_phi}. 
Again the observations are consistent with the existence of a phase transition, in this case at 0.6.
Since the bed yields for a force only 
about 30\% of the plate yield force, we conclude that the plate measurements are indicative of the yield shear
stress alone. 

\begin{figure}[t]
  \includegraphics*[width=8.5cm]{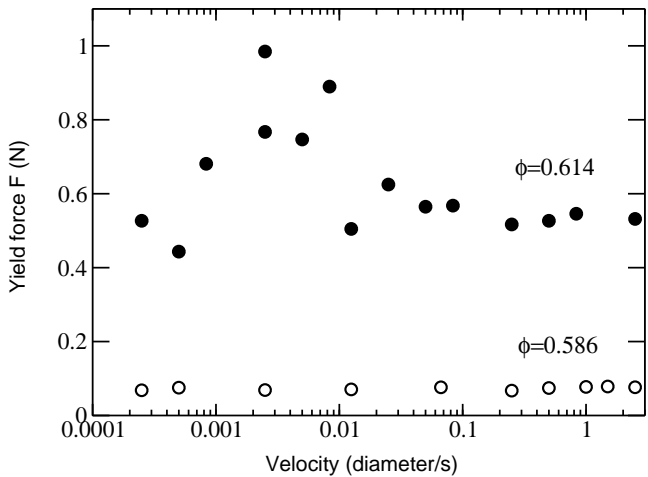}
  \caption{The yield force is velocity independent. All other measurement presented in this paper were made at a velocity of
0.0125 grain diameters per second.}
  \label{fig:yield_v}
\end{figure}

Figure \ref{fig:yield_v} shows that the measured yield force is independent of the velocity of the plate, both above and below
$\phi$ = 0.595. The larger fluctuations at lower velocities and higher packing fractions are an unexpected result.

\begin{figure}[t]
  \includegraphics*[width=8.5cm]{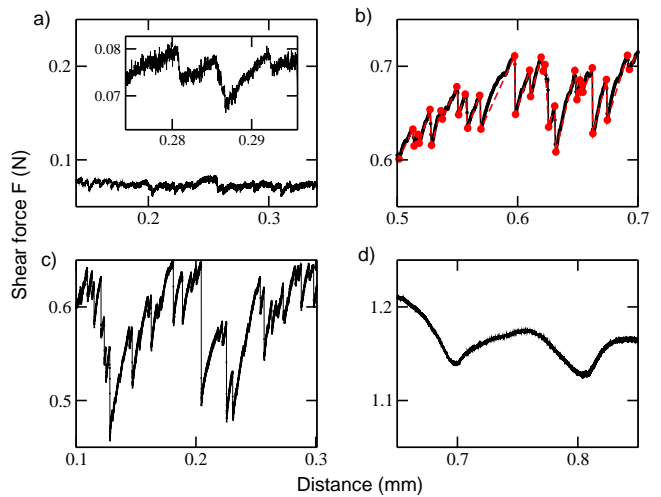}
  \caption{Stick-slip events are evident in force measurements for $\phi$ values (a) 0.584 (plate), (b) 0.623 (plate), and (c) 0.626 (rod), but not in the measurements in (d) for the plate at $\phi=0.626$.  
The dots in (b) are the minima and maxima used for the analysis; consecutive extrema are separated by at least 5 mN, which is about five times the noise level.}
  \label{fig:fluctuations}
\end{figure}

\begin{figure}[t]
  \includegraphics*[width=8.5cm]{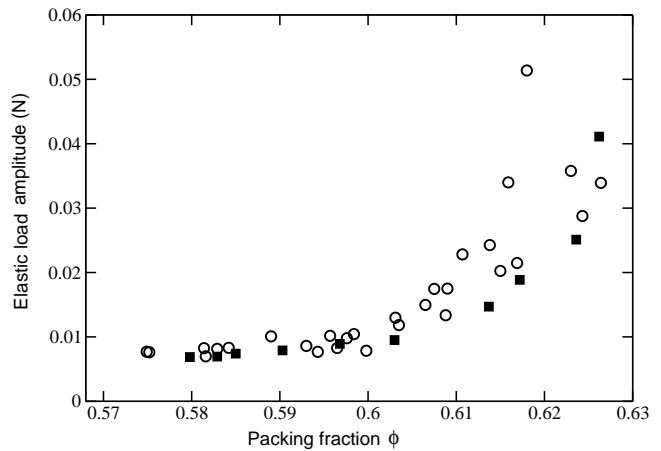}
  \caption{
The average $\Delta F$ during the elastic load phases. 
For $\phi > 0.6$,  the amplitudes measured with the plate (open circles) begin to become higher than
the amplitudes for the vertical rod  (filled squares). 
  }
  \label{fig:min_max}
\end{figure}

The fluctuations in the force measurements shown in Fig. \ref{fig:stress_displacement} contain information 
concerning the elastic response of the granular medium.  The force measurements reveal stick-slip motion, that is,
smooth ``elastic'' increases of force followed by rapid plastic decreases, as Fig. \ref{fig:fluctuations}(a)-(c) illustrates. 
An analysis of the fluctuations, inspired by work of Cantat and Pitois \cite{cantat:06}, yields an elastic constant
\begin{equation}
e = \frac{E \; A}{L} = \frac{\Delta F}{\Delta x},
\end{equation}
where $\Delta F/\Delta x$ is the slope of the elastic increases (cf. inset of Fig. 
\ref{fig:min_max}(a)), $A$ is the area of the plate, and $L$ is the $\it {a priori}$ unknown width of the sheared or compressed zone  
(cf. Fig. \ref{fig:setup}(b) and (c)). 
Since the rod mostly compresses the sample, its elastic modulus $E$ corresponds to the bulk modulus  $B$ of the sample.
For the plate, the measured force arises from both compression and shear, so $E$ is a combination of $B$
and the shear modulus $G$.

\begin{figure}[t]
  \includegraphics[width=7cm,trim=70 0 70 -37]{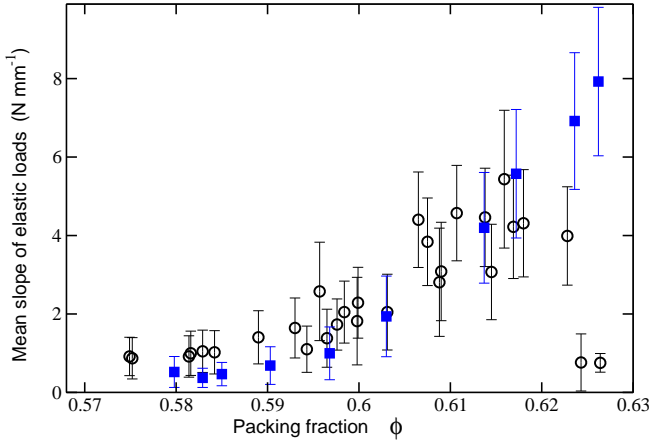}
  \caption{Elastic constants $e_p$ for the plate (open circles) and $e_r$ for the rod (filled squares).
Given the scatter, the data are consistent with a possible transition around $\phi = 0.595 $. 
The error bars correspond to the standard deviation of the $\Delta F/\Delta x$ slope distributions. 
} 
 \label{fig:g_phi}
\end{figure}

Figure  \ref{fig:g_phi} shows the $\phi$-dependence of the elastic constants $e_p$ for the plate and $e_r$ for the rod. The magnitudes of $e_p$ and $e_r$ are similar. Both elastic $e_p$ and $e_r$  exhibit a change in behavior at $\phi \approx 0.595$, just as found for the average elastic load amplitudes (Fig. \ref{fig:min_max}).
Additionally we found $e_p$  to be velocity-independent for a wide range of forcings, 
just as found for the shear force.  The values of $e_p$ in Fig.~\ref{fig:g_phi} obtained for the two highest 
packing fraction values are well below the line given by the nearby data; this is consistent with the absence of measured stick-slip motion in Fig.~\ref{fig:fluctuations}(d).

\section{Discussion}

We have found that both the yield force and its elastic response exhibit a change in behavior
at  $\phi \approx 0.595$, the same value of $\phi$  
(within experimental uncertainty) as the previously observed phase transition
in a penetration experiment  \cite{schroeter:07}. In contrast with the penetration experiment, 
the yield force measured here (Fig. \ref{fig:yield_phi}) 
is created mostly by shear. Recent measurements show that the onset of dilatancy  \cite{kabla:09,gravish:10} (the change in sign of the
volume response of a sheared system) also occurs for $\phi$ about 0.59-0.60, 
probably with small dependence on
friction coefficient.
The basic theoretical understanding of dilatancy goes back
to Reynolds \cite{reynolds:85}: a dense granular material must expand when sheared since
intertwined layers must separate if they are forced past one
another, while a low density granular material contracts rather than 
expands under shear. Whether the dilantancy {\it onset}, the transition
from contraction to expansion under shear, occurs so suddenly as to constitute a phase
transition has not been directly determined experimentally, but on the
theory side there is a model predicting this,
\cite{aristoff:10a}, an adaption of the Edwards model \cite{edwards:89} of granular
matter to include the above mechanism of Reynolds. 

Thus there are two shear responses, one associated with yield
exhibiting a transition (Fig.~\ref{fig:yield_phi}) 
and the other, dilatancy
onset, in which the Reynolds mechanism produces a phase transition in
a model.  Both responses are observed to occur at the same volume fraction
within the experimental uncertainty.  In addition, there is a phase transition
at the same volume fraction (again, within the experimental uncertainty) 
indicated by a response to penetration forces that are a mixture of shear and compression.
Since a  phase transition often leads to discontinuous
behavior in a variety of material properties, this suggests that the
yield transition data we report here, and
dilatancy onset, are aspects of the same phase transition.

\begin{figure}[t]
  \includegraphics*[width=5cm]{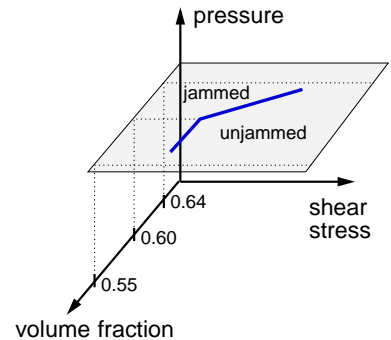}
  \caption{Sketch of the granular state diagram. 
    The two lines in the horizontal plane correspond to the exponential fits for the sheared plate in Fig.~\ref{fig:yield_phi}.
}
  \label{fig:phasediagram}
\end{figure}

A second main point concerns the interpretation of the shear yield
stress in Fig.~\ref{fig:yield_phi}. For small shear stress, the
system will flow but then arrest in a new configuration, while for
shear stress larger than the yield stress the system will flow and
not come to rest \cite{grebenkov:08}. 
These solid-like and liquid-like states can be
mapped in a diagram with the axes volume fraction, pressure, and shear stress,
as Fig.~\ref{fig:phasediagram} illustrates. 
Therefore, the data presented in Fig.~\ref{fig:yield_phi} correspond to the jammed-unjammed transition
in  the classical jamming state diagram \cite{liu:98,ohern:03}
at a given pressure \cite{zhang:10} and friction coefficient.

Jamming state diagrams in the literature are usually 
cartoons, but the separation of jammed and unjammed states could be made
quantitative by using experimental data, as in the use of the data from Fig.~\ref{fig:yield_phi} in Fig.~\ref{fig:phasediagram}. Diagrams for real 
granular matter might include other phase transitions, for example, a possible
transition at random close packing, $\phi \approx 0.64$ [13-15].
Including these transitions in a jamming state diagram would emphasize the need for a 
single theory that could simultaneously model both these 
phase transitions and the jamming/unjamming transition.

In connecting our results with the jamming state diagram for frictionless spheres,
we note some ambiguity in
the question of the equivalent of point J in frictional systems.
The jamming point $\phi_J = 0.64$ obtained in simulations of frictionless soft spheres corresponds to
two phenomena. First it is the only possible density of amorphous packings of undeformed pressure-free spheres.
Denser packings can only be built from compressed spheres and therefore need to have a  
finite pressure. This statement is independent of friction and hence also applicable
to a physical granular system, as shown in a  recent 2-dimensional experiment \cite{majmudar:07}. 
Second, point J is also the lower limit in which a system supports a positive yield stress. However, with friction 
there exist (as demonstrated by many experiments), for any pressure, mechanically stable states at a continuum 
of packing fractions below $\phi = 0.64$. 
The lowest volume fraction at which a finite yield stress can be measured then depends on pressure and
friction \cite{jerkins:08,farrell:10}.  Its zero pressure limit is the random loose packing (RLP) limit, and in recent work it has been repeatedly 
taken as the equivalent of point J \cite{zhang:05,shundyak:07,jerkins:08}. Figure \ref{fig:phasediagram} is based
on such an interpretation. 

Extrapolating from the penetration experiments in \cite{schroeter:07},
we expect a significant pressure dependence. 
Additionally, previous measurements have shown that, even at a constant depth in the
sample as in our measurements, the pressure depends on volume fraction and sample preparation
\cite{vanel:99}.  Therefore, future work should include independent
measurements of pressure. 

Our results on the elastic properties can be interpreted in the framework of a model in which bulk compression and shear 
are represented by compression and  stretching of parallel elastic springs. These springs can be thought of as the individual 
force chains originating at the probe.
In the plate measurements there are considerably more shear springs present than in the rod measurements.
The almost equal outcome for the two cases indicates that the shear elastic response is significantly weaker than 
the compression elastic response. 
If we assume that the relevant length scale $L$ is about the same for shear
and compression, this result is equivalent to assuming that the shear modulus of the granular medium is significantly smaller than its bulk modulus. 
In comparison, in the jamming paradigm the ratio of the shear modulus to the compression modulus goes to zero 
when the volume fraction approaches point J from above 
\cite{makse:99,ohern:03,ellenbroek:09}.

The velocity independence of both the yield shear stress and the elastic properties 
that we find contrasts with the previously observed logarithmic velocity 
dependence in sheared 2-dimensional photoelastic discs  \cite{hartley:03,behringer:08}.
This might be either related to the difference in dimensionality or to the
different values of the Young's moduli of the materials used:
70 GPa/m$^2$ for glass beads and 4 MPa/m$^2$ for photoelastic discs.

In this paper we have focused on unjamming properties under shear of 3-dimensional
packings of frictional spheres. We have found that both yield shear
stress and shear modulus per length exhibit a transition at packing
fraction $\phi\approx 0.595$. As we have discussed, this particular value 
of $\phi$ could correspond to a phase transition and could coincide with dilatancy onset.
In addition, we have shown that the shear stress is velocity independent 
over a range of four decades, both below and above $\phi\approx0.595$. 

{\bf Acknowledgments}
We acknowledge helpful discussions with Olivier Dauchot.
We would like to thank Udo Krafft for experimental support.
This research was supported in part by Welch Foundation Grant F-0805 and NSF Grant DMS-0700120.

\end{document}